\documentclass[twocolumn, noshowpacs, preprintnumbers, amsmath,amssymb,nopreprintnumbers]{revtex4}
\usepackage[colorlinks,urlcolor=blue]{hyperref}
\usepackage{dsfont}
\usepackage{graphicx}
\usepackage{lineno}
\usepackage{dcolumn}
\usepackage{bm}
\usepackage{appendix}
 \usepackage{verbatim}

\newcommand{\sign}{\mathrm{sign}\,}
\usepackage{breakurl}
\usepackage{amsmath}
\usepackage{enumerate}
\makeatletter
  \let\@font@info\@gobble
  \let\@font@warning\@gobble
\makeatother

\begin{document}

\title{Far tails of the biased CTRW model under the short time limit}
\author{Wanli Wang\email{wanliiwang@163.com}, Kaixin Zhang and Yuda Cheng}
\affiliation{School of Mathematical Sciences, Zhejiang University of Technology, Hangzhou 310023, China}

%



\date{\today}

\begin{abstract}
It has been observed in numerous experiments,  simulations, and various theoretical studies that the spreading of particles can be modeled by the continuous-time random walk. We consider two widely used cases, namely Gaussian and discrete displacements, to compute the position distribution and demonstrate the emergence of exponential decay in the far tails when a bias is introduced.
We further analyze the temporal rate function and the positional rate function to examine the convergence of the theoretical predictions. For Gaussian displacements, we also discuss the relationship between the position distributions with and without bias in different asymptotic limits.
\end{abstract}

%

%


\maketitle
The continuous-time random walk (CTRW) \cite{Hou2018Biased,Metzler2000random,Deng2020Modeling,Wang2020Fractional,Afek2023Colloquium,Montroll1965Random,Berkowitz2009Exploring,Alon2017Time}  extends the classical random walk by incorporating random waiting times between successive jumps. This concept was originally introduced by Montroll and Weiss \cite{Montroll1965Random}. In this framework, the evolution of the process is fully characterized by the probability distributions of jump lengths and waiting times. Different choices of the waiting time and displacement distributions give rise to distinct diffusion regimes \cite{Metzler2000random}, ranging from subdiffusion to normal diffusion to enhanced  diffusion.  When the process exhibits normal diffusion, one might instinctively expect the positional distribution to be Gaussian; however, this is not necessarily the case.  To be more exact, in Ref. \cite{Pinaki2007Universal}, Chaudhuri, Berthier, and Kob study the distribution of single particle displacements in a broad class of materials, for example, dense suspensions of colloidal hard spheres, where  they find that the tails follow exponential, rather than Gaussian. This behavior has been observed in many systems or experiments, including mRNA molecules in heterogeneous environments \cite{Thomas2017Cytoplasmic}, nano-particles in confined diffusion \cite{Xue2016Probing} and in a nanopost array \cite{Kai2013Diffusive}, a tracer particle in crowded media\cite{Surya2015Non-universal}, living cells \cite{Wang2009Anomalous}, stock markets \cite{Masolivera200dynamic} and diffusion of heterogeneous populations \cite{Hapca2009Anomalous}, the motion of chloroplasts in plant cells \cite{Schramma2023Chloroplasts}; see related works in Refs. \cite{Luo2019Quenched,Kege2000Direct,Weeks2000Three,Pinaki2007Universal,Wang2009Anomalous,Hapca2009Anomalous,Leptos2009Dynamics,Eisenmann2010Shear,Toyota2011Non,Skaug2013Intermittent,Xue2016Probing,Wang2017Three,Jeanneret2016Entrainment,Chechkin2017Brownian,Hu2023Triggering,Upadhyay2026Net}. 

In the context of the CTRW model, the Laplace tails are observed at the short time limit \cite{Barkai2020Packets}. When the mean number of renewals or jumps over the interval $(0,t)$ is very small (e.g., tending to zero), most particles remain trapped in their initial positions. Consequently, the central part of the positional distribution becomes less informative. Instead, the relevant limiting law appears in the statistics of tails. This regime is closely connected to large deviation theory \cite{Touchette2009large,Touchette2018Introduction,Adrian2021Large,Wang2020Large}, describing the rare events of the position.  As mentioned in \cite{Barkai2020Packets}, the exponential decay
is caused by many jumps moving in the same direction. For a short time $t$, many short waiting times between two successive
jumps are responsible for the mentioned exponential decay.

Substantial work has been devoted to the study of unbiased CTRWs.  Barkai and Burov found that exponential behavior is generally observed in a large class of problems of
transport provided that the PDF of the waiting time between two successive steps is analytic at small $\tau$ \cite{Barkai2020Packets}. The anti-bunching and bunching of jump events have been discussed to study large deviations of the position \cite{Wang2020Large}. Pacheco-Pozo and Sokolov investigated rate functions within the framework of large deviation theory for a broad class of waiting-time distributions, ranging from the exponential distribution to the one-sided L{\'e}vy distribution \cite{Adrian2021Large}. In the context of the CTRW model, super-exponential and sub-exponential displacements have been investigated using  large deviation theory and the single big jump principle \cite{Hamdi2024Laplace}.
Moreover, the exponential decay  was discussed in terms of diffusing diffusivity models \cite{Chechkin2017Brownian,Miotto2021Length}.

Heterogeneous media are ubiquitous in real systems, ranging from physical to biological environments \cite{Levy2003Measurement,Alon2017Time,Hou2018Biased}. Therefore, from a statistical perspective,  the study of biased dynamics is of considerable importance. In this manuscript, the far tails of the positional distribution are investigated within the framework of a biased CTRW model. One may naturally ask why this topic is of particular interest in statistical physics. To provide context, consider a system in which particle waiting times are drawn from a power-law distribution, while the jump lengths follow a Gaussian distribution with a nonzero mean. In the long-time limit, the positional distribution is governed primarily by the tail behavior of the waiting-time distribution. More precisely, if waiting times possess finite mean and variance, the resulting positional distribution is Gaussian \cite{Metzler2000random}. If waiting times have a finite mean but diverging variance, the positional distribution converges to an asymmetric L{\'e}vy stable distribution \cite{Alon2017Time,Wang2019Transport}. When the mean of waiting times diverges, the dynamics are different from the above scenarios and are instead described by the fractional time diffusion equation \cite{Metzler1999Anomalous}. However, the situation changes markedly in the short-time regime. In this limit, the influence of the heavy tails of the waiting-time distribution becomes negligible. Instead, the positional distribution is determined  by the behavior of the waiting-time density near the origin.

Biased CTRWs are widely used to describe diffusion in  heterogeneous medium. In this setting, bias fundamentally modifies the  mechanism of  trajectories and leads to qualitatively different far-tail statistics compared with the unbiased case \cite{Shlesinger1974Asymptotic,Gradenigo2016Field,Barkai2020Packets,Burov2022Exponential}. Understanding how bias reshapes the asymptotic behavior of the positional distribution is therefore essential for characterizing fluctuation properties in such systems. Note that our primary interest lies in the biased CTRW with a short-time limit. This was first investigated in Ref. \cite{Burov2022Exponential}, and its main findings are summarized as follows: (i) under the detailed balance assumption, the ratio between the right and left tails of the positional distribution grows linearly with position; (ii) the positional distribution with bias can be mapped to the scaled propagator without bias, revealing an exponential decay of the positional distribution. 
From a physical perspective, Ref.~\cite{Burov2022Exponential} provides a useful mapping-level description of biased CTRWs, but does not fully resolve the microscopic origin of the far-tail behavior. In this manuscript, we go beyond this description and investigate the renewal-level mechanism underlying the positional distribution. A central question is which parameters control the exponential decay of the positional distribution, as this provides  a microscopic interpretation of its universality. A second objective is to develop a practical way to quantify the role of the bias using the statistics of the tails, in particular, to provide a simple estimator in the weak-bias regime. 
The rate functions, as key objects in both physics and mathematics, are well understood for independent and identically distributed (IID) random variables \cite{Touchette2018Introduction}. However, in the present context, the waiting times are not independent; instead, they are correlated through the structure
$\tau_1 + \tau_2 + \cdots + \tau_N + B_t$,
where $B_t$ denotes the backward recurrence time \cite{Godreche2001Statistics}. Consequently, the rate function encodes genuinely dynamical information beyond a simple reformulation of the positional distribution. Finally, motivated by discrete transport in physical systems such as vacancy and interstitial diffusion in solids, the case of discrete jump lengths—and how discreteness modifies far-tail behavior—would be an interesting direction  to explore.

In this manuscript, we obtain the far tails of the positional distribution within the framework of a biased CTRW model, using both a Gaussian jump distribution and a two-point (binary) distribution. Our results demonstrate that a nearly exponential decay emerges for the far tails.
A key result is that the dominant contribution to the exponential behavior is governed by the most likely number of renewals $N^*$ required to reach a given position. This quantity controls the scaling form of the positional distribution and depends on microscopic model details. This point was not addressed in Ref.~\cite{Burov2022Exponential}. We further propose a novel way to quantify the effect of bias by comparing the tails of the positional distribution with and without bias. In particular, in the weak-bias regime, we derive a simple linear relation that characterizes this effect. The mapping method of Ref.~\cite{Burov2022Exponential}, however, is not well suited for extracting the leading and subleading bias terms. We also analyze the corresponding rate functions within the framework of large deviation theory, which are important observables in both physical and mathematical fields. The rate functions are obtained analytically for the considered cases and are verified by numerical simulations. In contrast to much of the existing literature, which primarily focuses on long-time asymptotics, the present work emphasizes the short-time regime, which is particularly relevant for transient transport and non-equilibrium fluctuations. Overall, the present results provide a clearer understanding of the exponential decay structure in biased CTRWs. We note that  CTRW models are widely used in contexts such as transport on networks, chemical tracer dynamics \cite{Alon2017Time}, and diffusion in complex media \cite{Hu2023Triggering}. The theoretical predictions developed here may be relevant to experimental observations and data analysis.

This manuscript is organized as follows.
In Sect. \ref{AT19SEC2} we introduce the CTRW model.  The far tails of the distribution and the rate functions are investigated in Sect.~\ref{AT19SEC3} using the large deviation theorem for Gaussian displacements.  In Sect. \ref{brwalk}, we consider the binomial random walk and the far tails of the positional distribution. Finally, we conclude the manuscript.

\section{Model}\label{AT19SEC2}
In the CTRW model \cite{Hou2018Biased,Metzler2000random,Deng2020Modeling,Wang2020Fractional,Montroll1965Random,Berkowitz2009Exploring,Alon2017Time,Afek2023Colloquium}, the position of the random walker, $x(t)=\sum_{i=1}^N x_i$, is determined by the distribution of waiting times and displacements.
Here, $N$ is a time-dependent random integer representing the number of jumps occurring from time $0$ to time $t$. The particle is initially located at the origin
at time $t=0$ and waits at its initial position for a random time $\tau_1$ drawn from a PDF $\phi(\tau)$. After that, the particle immediately undergoes a displacement  $x_1$ with PDF $f(x)$. At time $t=\tau_1$, we generate a new waiting time $\tau_2$ and a  displacement $x_2$ according to the corresponding PDFs $\phi(\tau)$ and $f(x)$, respectively. Then, the process is repeated. Here, the waiting time $\tau_i$ and the displacement $x_i$ are IID random variables.
Throughout, $x_0$ denotes the particle's initial position; in our setting, $x_0=0$.

We denote $f(x|N)$ \cite{Metzler2000random} as the probability of finding the particle at position $x$, given that it has made $N$ jumps. $Q_t(N)$ corresponds to the probability distribution of the number of events between zero and $t$ \cite{Godreche2001Statistics}.
Let $P(x,t)$ be the PDF of finding the particle at position $x$ at time $t$, which can be expressed exactly as an infinite series
\begin{equation}\label{AT19SEC2EQ1201}
\begin{split}
  P(x,t) &=\sum_{N=0}^\infty  Q_t(N)f(x|N) \\
    &\rightarrow \int_0^\infty Q_t(N)f(x|N)dN,
\end{split}
\end{equation}
where  $N$ is treated as a continuous variable. The second line of Eq.~\eqref{AT19SEC2EQ1201}, corresponding to the so-called integral formula
of subordination, is valid in the limit of large $N$. Assuming the displacements of the particle are IID random variables, so $
\widetilde{f}(k|N)=\widetilde{f}^N(k)$ with  $\widetilde{f}(k)=\int_{-\infty}^{\infty}\exp(ikx)f(x)dx$ being the Fourier transform of $f(x)$.
In Laplace space, the probability distribution of the number of events between $0$ and $t$ reads \cite{Godreche2001Statistics}
\begin{equation}\label{AT19SEC2EQ1202}
 \widehat{Q}_s(N)=\frac{1-\widehat{\phi}(s)}{s}\widehat{\phi}^N(s),
\end{equation}
where $\widehat{\phi}(s)=\int_0^\infty\phi(\tau)\exp(-s\tau)d\tau$ denotes the Laplace transform of $\phi(t)$, from the real space $t$ to the Laplace space $s$.
In the particular case of $N=0$,
we have $Q_t(0)=\int_t^\infty \phi(\tau)d\tau$, which corresponds to the survival probability.

Recall that we are interested in the positional distribution in the short-time limit, so the short-time statistics of the number of renewals are important. Here, we follow the approach of \cite{Barkai2020Packets}, where $\phi(\tau)$ is assumed to be analytic at the critical point $\tau=0$. Mathematically, we assume that $\phi(\tau)$ obeys the following Taylor expansion \cite{Barkai2020Packets}
\begin{equation}\label{AT19SEC2EQ101}
  \phi(\tau)\sim  \sum_{j=0}^\infty C_{A+j}\tau^{A+j},~~~\tau\to 0,
\end{equation}
where $A\geq 0$ is a non-negative integer. Note that the parameter $A$ can be extended to the case when $A>-1$. For further details, see Ref.~\cite{Wang2024Statistics}. The advantage of Eq.~\eqref{AT19SEC2EQ101} is that the theory developed below applies to a broad class of waiting-time distributions, ranging from exponential to power-law forms. 
For large
$N$, using Eqs.~\eqref{AT19SEC2EQ1202} and \eqref{AT19SEC2EQ101}, the far tail of renewals distribution behaves as \cite{Barkai2020Packets}
\begin{equation}\label{StasEli}
Q_t(N)\sim \frac{\exp\left(\frac{C_{A+1}}{C_{A}}t+N(A+1)\ln\left(\frac{tze}{N}\right)\right)}{\sqrt{2\pi N(A+1)}}
\end{equation}
with
\begin{equation}\label{AT19SEC3102aty}
  z=\frac{[C_A\Gamma(A+1)]^{\frac{1}{1+A}}}{(A+1)}.
\end{equation}
Note that the leading term, corresponding to the bracketed expression in Eq.~\eqref{StasEli}, is $N(A+1)\ln(tze/N)$, indicating that the far tail of the distribution of $N$ exhibits a nearly exponential decay.

For the displacements, two widely applicable cases are considered for theoretical purposes: the Gaussian distribution in the continuous setting and the two-point distribution in the discrete setting \cite{Klafter2011First}. The case of Gaussian displacements with a nonzero mean is a well-studied paradigm and an interesting topic in statistical physics, together with power-law waiting times. The discrete case, on the other hand, is relevant not only to anomalous diffusion but also to network-related processes \cite{Upadhyay2026Net}.
Both types of displacements possess finite mean and variance, which play a crucial role in determining the exponential decay of the positional distribution.

Below, Eq.~\eqref{StasEli} will be employed as a key analytical tool to investigate the far-tail behavior of the positional distribution. To obtain an analytical prediction, the saddle-point method is applied to the second line of Eq.~\eqref{AT19SEC2EQ1201} in the limit of large $N$. The aim of the manuscript is to provide further evidence for the exponential decay and present a more
detailed analysis of the model covering cases not discussed previously.

\section{Exponential tails of the  position distribution with Gaussian displacements}\label{AT19SEC3}

We begin with a scenario where the PDF of the waiting time exhibits analytic behavior as $\tau$ approaches zero and the step length follows a Gaussian distribution. Subsequently, we analyze the far tails of the positional distribution, explore the Einstein-like relation,  and examine the rate function.

\subsection{Far tails of the positional distribution}

For the displacement, we focus on a Gaussian distribution with  mean $a>0$ and variance $\sigma^2$
\begin{equation}\label{AT19SEC2EQs104}
  f(x)=\frac{1}{\sqrt{2\pi\sigma^2}}\exp\left[-\frac{(x-a)^2}{2\sigma^2}\right].
\end{equation}
As noted earlier, we assume the displacements $x$ are IID random variables. For a given $N$, the probability density $f(x|N)$ is given by
\begin{equation}\label{AT19SEC2EQs105}
f(x|N)=\frac{1}{\sqrt{2\pi\sigma^2 N}}\exp\left(-\frac{(x-aN)^2}{2\sigma^2N}\right).
\end{equation}
According to Eq.~\eqref{AT19SEC2EQ1201}, we have
\begin{widetext}
\begin{equation}\label{AT19SEC3100}
 P(x,t)\sim\int_0^\infty  \exp\left(-\underbrace{N\left[\left(\frac{x/N-a}{\sqrt{2\sigma^2}}\right)^2-(A+1)\ln\left(\frac{zte}{N}\right)-\frac{C_{A+1}}{C_A}\frac{t}{N}+\frac{\ln (2\pi\sigma^2 N\sqrt{A+1})}{N} \right]}_{\chi(N)}\right)dN.
\end{equation}
\end{widetext}
For Eq.~\eqref{AT19SEC3100}, we used the large deviation result for the number of renewals \cite{Barkai2020Packets}, namely Eq.~\eqref{StasEli}.
In turn, this means that Eq.~\eqref{AT19SEC3100} is valid for large $x$ since the mean and variance of displacements are finite.  Generally, it is not easy to perform the integral to get an analytic solution.
Here we turn to the
Cram{\'e}r-Daniels approach \cite{Daniels1954Saddlepoint} (sometimes called the saddle point approximation), and apply it to our problem. For large $|x|$, from the saddle point method and Eq.~\eqref{AT19SEC3100}, the far tails of the positional distribution are
\begin{equation}\label{saddlePointM}
P(x,t)\sim \frac{\sqrt{2\pi}}{\sqrt{|\chi^{''}(N^*)|}}\exp(-\chi(N^*)),
\end{equation}
where the saddle point $N^*$  satisfies $\chi^{'}(N^*)=0$, namely
\begin{equation}\label{AT19SEC3fsdG101}
\frac{a^2}{\sigma ^2}-2 (A+1) \ln \left(\frac{t z}{N^{*}}\right)-\frac{x^2}{(N^*)^2 \sigma ^2}+\frac{1}{N^*}=0.
\end{equation}
In the  large $N^{*}$ limit, the asymptotic solution of Eq.~\eqref{AT19SEC3fsdG101} reads
\begin{equation}\label{AT19SEC3101}
 N^*\sim\frac{|x|}{\sigma\sqrt{(A+1)W_0\left[\frac{(A+1)   \exp\left(\frac{a^2}{(A+1) \sigma ^2}\right)}{(C_A\Gamma (A+1))^{\frac{2}{A+1}} \sigma ^2}\frac{x^2}{t^2}   \right]}},
\end{equation}
where $W_0(y)$ is the principal branch of the Lambert $W$ function \cite{CorlessOn1996,Zarfaty2016Statistics,Zarfaty2018Dispersion}, also called the product logarithm function.
When $|x|\to\infty$, Eq.~\eqref{AT19SEC3101} reduces to
\begin{equation}\label{AT19SEC3fsdG102}
N^*\propto   \frac{ |x|}{\sigma\sqrt{(A+1)\ln(\exp( \frac{a^2}{(A+1)\sigma^2})\frac{x^2}{t^2})}}.
\end{equation}
An interesting feature of $N^{*}$ is exclusively
exhibited by the large deviation analysis, i.e., when the force increases, $N$ decreases. This phenomenon can be explained as follows: When the bias is strong, the particle requires fewer steps to reach a fixed position than when the bias is weak. Let $N^-$ and $N^+$ be the  most likely number of steps of the particle reaching the site $-|x|$ and $|x|$, respectively.
Clearly, $N^+$ and $N^-$ depend on  $a$ and $\sigma$. When $|x|\gg \langle x(t)\rangle$, position-dependent  $N^+$ and $N^-$ are the same based on Eq.~\eqref{AT19SEC3fsdG102}.  For a fixed $a$ and $x\to\infty$, Eq.~\eqref{AT19SEC3fsdG102} yields
\begin{equation}
\frac{N^*}{|x|}\sqrt{\ln\left(\frac{|x|}{t}\right)}\propto \frac{1}{\sigma\sqrt{2(A+1)}},
\end{equation}
tending to a constant.

\begin{figure}[htb]
 \centering
 \includegraphics[width=0.5\textwidth]{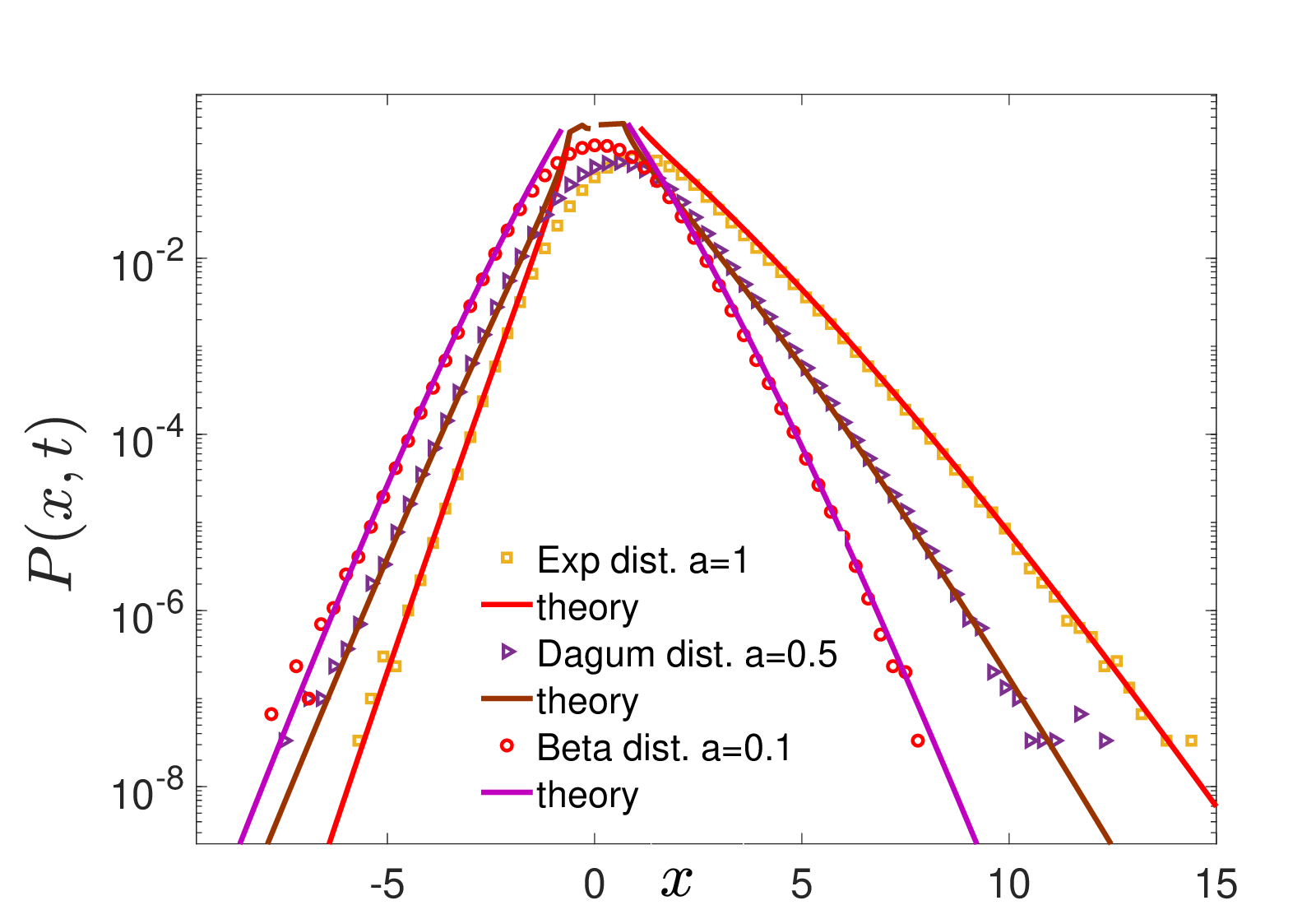}\\
 \caption{The positional distribution exhibits asymmetric exponential tails for different biases, considering various waiting-time PDFs, such as exponential distribution $\phi(\tau)=\exp(-\tau)$ (`$\square$'), Dagum distribution $\phi(\tau)=1/(1+\tau)^2$ (`$\rhd$') and a special form of Beta distribution $\phi(\tau)=6\tau(1-\tau)$ (`$\circ$').
The symbols represent simulations generated from $10^9$ trajectories and  the solid lines depict theoretical predictions based on Eq.~\eqref{AT19SEC3fsds101}.  Note that, in all cases considered here, non-moving particles are not included.
In our setting, we choose $\sigma=1$ and $t=0.5$ for different values of $a$.
}\label{PxtNew}
\end{figure}

According to Eq.~\eqref{saddlePointM}, we find the main result of this section
\begin{equation}\label{AT19SEC3fsds101}
\begin{split}
  P(x,t) &\sim \frac{1}{\sqrt{2\pi\sigma (A+1)^{3/2} }}\\
  &\times\frac{\exp \left(\frac{a }{\sigma ^2}x+\frac{C_{A+1} }{C_A}t-|x|G(\frac{x}{t})\right)}{  \sqrt{\left||x|\left(\sqrt{W_0(\frac{gx^2}{t^2})}+\frac{1}{\sqrt{W_0(\frac{gx^2}{t^2})}}\right)-\sigma\right|}}
\end{split}
\end{equation}
with
\begin{equation*}
  \begin{split}
   G\left(l\right) & =\frac{\sqrt{A+1}}{\sigma}\left(\sqrt{W_0\left(gl^2\right)}-\frac{1}{\sqrt{W_0\left(gl^2\right)}}   \right)
  \end{split}
\end{equation*}
and
\begin{equation}\label{AT19SEC3102}
  g=\frac{\exp(\frac{a^2}{(1+A)\sigma^2})}{(1+A)\sigma^2z^2}.
\end{equation}
See Fig.~\ref{PxtNew} for a comparison with the simulation results. Here the observation time $t$ is expressed in dimensionless units by rescaling with a characteristic time scale $\tau_0$ of the underlying stochastic process. For simplicity, we set $\tau_0=1$. Therefore, $t=1/2$ represents a fraction of this characteristic time scale rather than a quantity with explicit physical units.

When a bias is added, the two exponential tails become asymmetric. Specifically, the right tail is $\exp(2ax/\sigma^2)$ times larger than the left if $a>0$.
Note that there is a term $a/\sigma^2 x$ in the numerator of Eq.~\eqref{AT19SEC3fsds101}, which indicates that one of the tails decays rapidly if the bias is strong.

Based on Eq.~\eqref{AT19SEC3fsds101}, the asymptotic behavior of the positional distribution is
\begin{equation}\label{AT19SEC3fsds10143}
\begin{split}
  P(x,t) &\propto \exp \left(-x \text{sign}(x)G(l)+\frac{a }{\sigma ^2}x+\frac{C_{A+1}}{C_A}\right).
\end{split}
\end{equation}
It demonstrates that the far tails of the  position distribution decay like an exponential function instead of a Gaussian distribution since $W_0(|y|)$ increases as $\ln(|y|)$ for large $|y|$.
What we would like to mention is that the statistics of the tails of waiting-time PDFs are not essential and the tails of the positional distribution are simply determined by the Taylor expansion at small $\tau$.

\subsection{Einstein-like relation}\label{AT19SEC3fsisd100}
\begin{figure}[htb]
 \centering
 \includegraphics[width=0.5\textwidth]{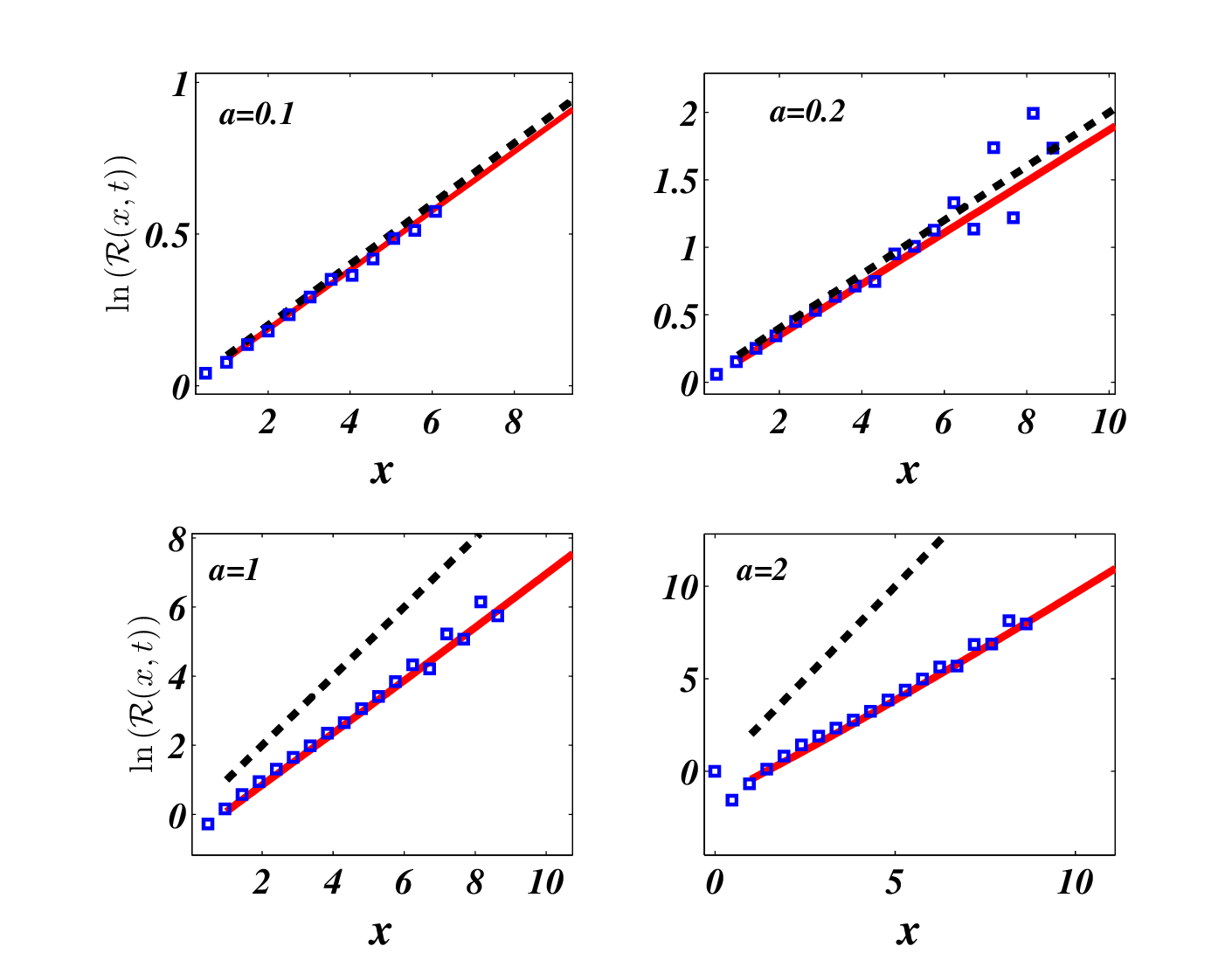}\\
 \caption{Plot of $\ln(\mathcal{R}(x,t))$ versus  the position $x$ for  different variables $a$ where waiting times are drawn from Dagum distribution $\phi(\tau)=1/(1+\tau)^2$.
 The red solid lines correspond to the theoretical result given by Eq.~\eqref{AT19SEC3fsis1001}, which is valid for all types of biases.  The linear relation, given by Eq.~\eqref{AT19SEC3fsis1004}, is represented by the dashed lines. The linear prediction (shown by the dashed line) is derived under the assumption of small bias, i.e., small \(a\). The discrepancies observed at larger \(a\) originate from the breakdown of this approximation, as Eq.~\eqref{AT19SEC3fsis1004} is no longer valid beyond the weak-bias regime. The symbols are simulations generated from $10^9$ realizations with $\sigma=1$.
The figure shows that as we decrease $a$, the linear relationship Eq.~\eqref{AT19SEC3fsis1004} becomes readily detected.  
%
%
}\label{FarTailPlusForceAndFreeDagumGaussian}
\end{figure}

We delve deeper into the far tails of the positional distribution and explore the impact of the bias. Inspired by the Einstein relation \cite{Barkai2003Aging,Klafter2011First,Wang2017Aging,Gradenigo2012Einstein} describing the relation between the mean of the position in the presence of bias and the variance of the position in the absence of bias, our aim is to investigate the ratio of the positional PDF with and without bias.  For simplicity, we introduce the symbol $\mathcal{R}(x,t) = P(x,t)_{a \neq 0}/P(x,t)_{a = 0}$ to represent the ratio of the tails. 
Based on Eq.~\eqref{AT19SEC3fsds101}, we have
\begin{equation}\label{AT19SEC3fsis1001}
\begin{split}
\mathcal{R}(x,t)&\sim \exp\Big(\frac{a}{\sigma^2}x-x \text{sign}(x)\\
 &~~~\times  \Big[G\Big(\frac{x}{t}\Big)_{a\neq 0}-G\left(\frac{x}{t}\right)_{a=0}\Big]\Big).
\end{split}
\end{equation}
Note that Eq.~\eqref{AT19SEC3fsis1001} is valid for large $|x|$.
Rewriting the above equation,  we find
\begin{equation}\label{AT19SEC3fsis1001add1}
\begin{split}
\ln\left(\mathcal{R}(x,t)\right)&\sim \frac{a}{\sigma^2}x-x \text{sign}(x)\\
&~~~\times\left[G\left(\frac{x}{t}\right)_{a\neq 0}-G\left(\frac{x}{t}\right)_{a=0}\right],
\end{split}
\end{equation}
which is shown as the red solid lines in Fig.~\ref{FarTailPlusForceAndFreeDagumGaussian}. An interesting limit involving $a$ could have significant practical applications.
Recall that $G(x/t)$ is an even function with respect to bias $a$. Therefore, the Taylor series of $G(x/t)$ with respect to small $a$ can be expressed as follows
\begin{equation}\label{AT19SEC3fsis1002}
G\left(\frac{x}{t}\right)=G\left(\frac{x}{t}\right)_{a=0}+\frac{1}{2!}\frac{d^2 G\left(\frac{x}{t}\right)}{d a^2}a^2+\cdots.
\end{equation}
From Eq.~\eqref{AT19SEC3fsis1002}, it follows that for a small bias, the right-hand side of Eq.~\eqref{AT19SEC3fsis1001add1} grows approximately linearly with $x$.
It indicates that the leading term of the right-hand side of Eq.~\eqref{AT19SEC3fsis1001} for small $a$ is
\begin{equation}\label{AT19SEC3fsis1003}
\mathcal{R}(x,t)\sim \exp\left(\frac{a}{\sigma^2}x\right)
\end{equation}
or
\begin{equation}\label{AT19SEC3fsis1004}
 \ln\left(\mathcal{R}(x,t)\right)\sim \frac{a}{\sigma^2}x;
\end{equation}
see the dashed line in Fig.~\ref{FarTailPlusForceAndFreeDagumGaussian}.
When the parameter $a$ increases, correction terms in Eq.~\eqref{AT19SEC3fsis1002}, such as $a^2$ and $a^4$, come into play. It indicates that Eqs.~\eqref{AT19SEC3fsis1003} and \eqref{AT19SEC3fsis1004} become invalid. To address this, we still assume that the bias is weak, but a correction term is added  to Eq.~\eqref{AT19SEC3fsis1003}. From Eq.~\eqref{AT19SEC3fsis1001}, we obtain
a useful expression
\begin{widetext}
\begin{equation}\label{AT19SEC3fsis1005}
\mathcal{R}(x,t)\sim \exp\left[\frac{a x}{\sigma ^2}-\frac{\sqrt{A+1} \left| x\right|} {\sigma } \sqrt{W_0\left(\frac{\left(\frac{x}{t}\right)^2}{(A+1) \sigma ^2 z^2}\right)} \left(\sqrt{\frac{a^2}{\left((A+1) \sigma ^2\right)W_0 \left(\frac{\left(\frac{x}{t}\right)^2}{(A+1) \sigma ^2 z^2}\right)}+1}-1\right)\right].
\end{equation}
\end{widetext}
If we replace the terms in the last parentheses by their equivalent values, then in the limit of large $|x|$, Eq.~\eqref{AT19SEC3fsis1005} reduces to:
\begin{equation}\label{AT19SEC3fss1006}
 \ln\left(\mathcal{R}(x,t)\right)\sim \frac{a x}{\sigma ^2}-\frac{\frac{a^2  \left| x\right| }{ 2 \sigma ^3\sqrt{A+1}}}{\sqrt{W_0\left(\frac{x^2}{(A+1) \sigma ^2 z^2t^2}\right)}}.
\end{equation}
The last term on the right-hand side of Eq.~\eqref{AT19SEC3fss1006} approaches a constant as $|x|$ becomes small. For small $x$, the function $W_0(|x|)$ can be approximated as $|x|$. If $|x|\to \infty$, the second term grows as $|x|/\ln(|x|)$, becoming significant for large $|x|$. This explains why the dominant term is effective for weak bias.
See the dotted line in  Fig.~\ref{FarTailPlusForceAndFreeDagumGaussian}.

\subsection{Rate function}\label{AT19SEC3fewsis100}
The central limit theorem states that when independent random variables are added, their properly normalized sum approaches a normal distribution as the sample size increases. Instead of focusing on the central part of the distribution described by the central limit theorem, we focus on the statistical behavior of the far tails using large deviation theory.
The rate function \cite{Touchette2009large,Daniel2018Anomalous,Touchette2018Introduction} is known as the Cram{\'e}r function and is sometimes referred to as the entropy function, which is widely used to quantify the probabilities of rare events. In this context, we consider two types of rate functions, one associated with the position and the other with the time. While the main idea remains the same for both approaches, the distinction lies in the limiting procedure, namely $x\to \infty$ or $t\to\infty$.

\begin{figure}[htb]
 \centering
 \includegraphics[width=0.5\textwidth]{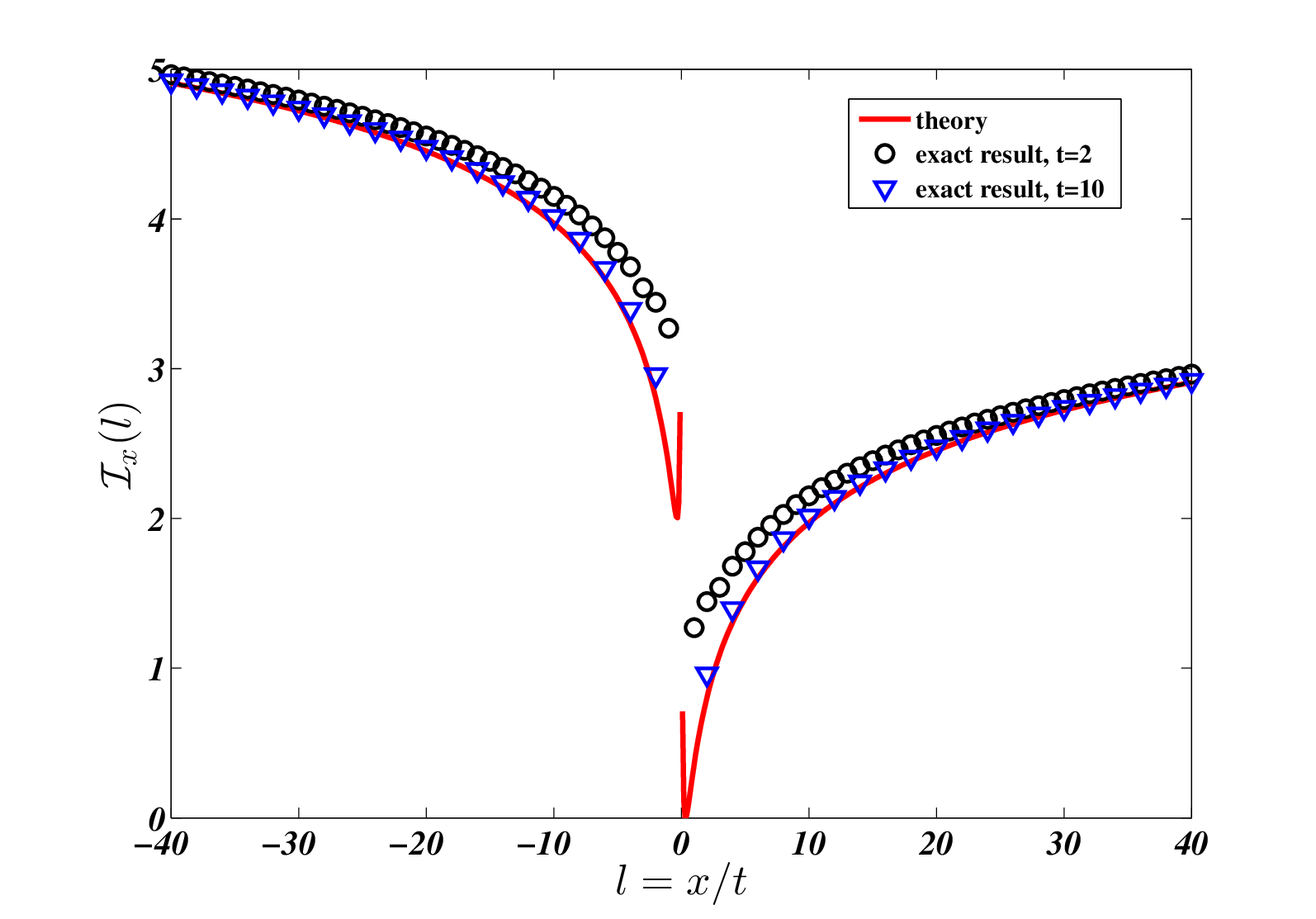}\\
 \caption{Rate function $\mathcal{I}_x(l)$ plotted for different times $t$. Here, the waiting time follows the Erlang distribution $\phi(\tau)=\tau^2\exp(-\tau)/2$ with mean $3$. In that sense $A=2$, $C_2=1/2$ and $C_{3}=-C_2$.
 The red solid line is the theoretical prediction Eq.~\eqref{AT19SEC3fsds102} without fitting. The exact results plot $\ln(P(x,t)/(-|x|))$, where $P(x,t)$ is estimated from Eq.~\eqref{AT19SEC2EQ1201} using the infinite terms \cite{Wang2020Large}. In our setting $a=1$ and $\sigma=1$. Though the observation time $t$ is not very long,  the convergence to the exact result is fast.  Note that as \(l \to 0\), our approximation diverges and, as expected, breaks down.
}\label{RateFunctionX}
\end{figure}

Utilizing Eq.~\eqref{AT19SEC3fsds101}, we have
\begin{equation}\label{AT19SEC3fewsis1001}
P(x,t)\sim \exp\left(-|x|G\left(\frac{x}{t}\right)+\frac{a }{\sigma ^2}x+\frac{C_{A+1} }{C_A}t\right).
\end{equation}
Rewriting Eq.~\eqref{AT19SEC3fewsis1001}, the rate function with respect to $x$ is
\begin{equation}\label{AT19SEC3fsds102}
 \lim_{t\to \infty}\frac{\ln(P(x,t))}{-|x|}=\mathcal{I}_x\left(l=\frac{x}{t}\right)
\end{equation}
with
\begin{equation}\label{AT19SEC3fewsis1001add11}
\mathcal{I}_x(l)=G(l)-\frac{a}{\sigma^2}\sign(l)-\frac{C_{A+1}}{C_A}\frac{1}{|l|}.
\end{equation}
Note that for large $l$ the asymptotic behavior of Eq.~\eqref{AT19SEC3fewsis1001add11} is
\begin{equation}\label{AT19SEC3fewsis1001add12}
\mathcal{I}_x(l)\sim\frac{\sqrt{A+1}}{\sigma}\sqrt{\ln(gl^2)},
\end{equation}
which is responsible for the nearly exponential decay of the positional distribution. As shown in Fig.~\ref{RateFunctionX}, the rate function $\mathcal{ I}_x(l)$ goes to infinity when $l\to 0$. On the other hand, when $l\to \infty$, Eq.~\eqref{AT19SEC3fewsis1001add12} also approaches infinity, but at an extremely slow rate.

Similarly, the (time) rate function follows
\begin{equation}\label{AT19SEC3fsds103}
 \lim_{t\to \infty}\frac{\ln(P(x,t))}{-t}=\mathcal{I}_t\left(l=\frac{x}{t}\right)
\end{equation}
with $\mathcal{I}_t(l)=\sign(l)l\mathcal{I}_x(l)$, where we obtained the time rate function using the positional rate function Eq.~\eqref{AT19SEC3fsds102}  without losing of information.

\section{Binomial random walk}\label{brwalk}
 In the previous section, we considered the case of continuous displacements and below we turn to the discrete case, i.e., a two-point distribution. Note that the waiting-time distributions considered here still follow the assumption given in Eq.~\eqref{AT19SEC2EQ101}. More precisely, we first analyze the case of exponential waiting times, and then turn to the more general case described by Eq.~\eqref{AT19SEC2EQ101}.

We consider a one-dimensional lattice random walk, starting from the origin. Suppose a step to the right occurs with probability $p$, while a leftward step has  probability $q=1-p>0$.
The particle is restricted to walk on the sites $j=\cdots, -3,-2,-1,0,1,2,3,\cdots$. Let the step length be one.
This concept was first introduced by Karl Pearson in a letter to Nature \cite{Pearson1905problem}.
Mathematically, the displacement follows a two-point distribution
\begin{equation}\label{AT19APP4eq102}
 f(x)=\left\{
               \begin{array}{ll}
                 1, & \hbox{probability $p$ to the right;} \\
                 -1, & \hbox{probability $q=1-p$ to the left}
               \end{array}
             \right.
\end{equation}
with $0< p<1$. In particular, if $p=1/2$, it means that the particles have an equal probability of jumping left and right, i.e., the symmetric case. 

\subsection{Exponential waiting time PDF}
Let us start with the exponential distribution given in Eq.~\eqref{expPDF} with unit mean. From Eq.~\eqref{AT19SEC2EQ1201}, for IID displacements, we have
\begin{equation}\label{ATAPPERW1001a1}
\begin{split}
 \widetilde{P}(k,t)
    & =\sum_{N=0}^\infty Q_t(N)\widetilde{f}(k)^N.
\end{split}
\end{equation}
Recall that the PDF of the number of renewals follows a Poisson distribution, i.e., $Q_t(N)=\exp(-t)t^N/N!$. Summing the infinite terms,  Eq.~\eqref{ATAPPERW1001a1} yields
\begin{equation}\label{adlsew101}
\begin{split}
 \widetilde{P}(k,t)&=\exp(t(\widetilde{f}(k)-1))\\
 &=\exp(tp\exp(ik)+tq\exp(-ik)-t).
\end{split}
\end{equation}
We further introduce the moment-generating function, which is the expectation of a function of the random variable. Mathematically, it can be written as $K(u)=\langle \exp(ux)\rangle=\int_{-\infty}^\infty P(x,t)\exp(ux)dx$. Setting $k=-iu$, from Eq.~\eqref{adlsew101} the moment generating function reads
\begin{equation}\label{adlsew102}
K(u)=tp\exp(u)+tq\exp(-u)-t.
\end{equation}
Instead of performing the inverse Fourier transform on Eq.~\eqref{adlsew101}, the saddle point approximation yields the large deviation expression \cite{Daniels1954Saddlepoint}
\begin{equation}\label{ATAPPERW1001a3}
P(x,t)\sim \frac{1}{\sqrt{2\pi K^{''}(u^{*})}}\exp(K(u^*)-u^{*}x),
\end{equation}
where $u^{*}$ is the solution of $K^{'}(u^{*})=x$. 
Based on Eq.~\eqref{adlsew102}, the solution of $K^{'}(u^{*})=x$ reads
\begin{equation}\label{ATAPPERW1001a5}
 u^{*}=\ln \left(\frac{x+ \sqrt{4 p q t^2+x^2}}{2 p t}\right).
\end{equation}
The other root $u^{*}$,
\begin{equation}\label{ATAPPERW1001a5as}
 u^{*}=\ln \left(\frac{x- \sqrt{4 p q t^2+x^2}}{2 p t}\right),
\end{equation}
does not satisfy the original equation.
In addition, from Eqs.~\eqref{adlsew102} and \eqref{ATAPPERW1001a5}, we get
\begin{equation}\label{ATAPPERW1001a6}
K^{''}(u^{*})=\sqrt{4 p q t^2+x^2}\propto |x|.
\end{equation}
Utilizing Eqs.~\eqref{ATAPPERW1001a3},  \eqref{ATAPPERW1001a5}, and \eqref{ATAPPERW1001a6}, we find the far tails of the positional distribution
\begin{equation}\label{AT19APP1eqbwe3001}
\begin{split}
P(x,t)\sim& \frac{\exp (-x \ln (\frac{\sqrt{4 p q t^2+x^2}+x}{2 p te})+\frac{4 p q t^2}{\sqrt{4 p q t^2+x^2}+x}-t)}{ \sqrt{2 \pi|x|}}.
\end{split}
\end{equation}
The corresponding simulation results are presented in Fig.~\ref{ExpPxt} of Appendix~\ref{CCCCC}.
When $p=q=1/2$, Eq.~\eqref{AT19APP1eqbwe3001} reduces to the symmetric case \cite{Wang2020Large}. Taking $x\to\infty$, the asymptotic solution follows
$$P(x,t)\sim\exp \left(-x \ln \left(\frac{\sqrt{4 p q t^2+x^2}+x}{2 p te}\right)\right),$$
showing the nearly exponential decay.

Rewriting Eq.~\eqref{AT19APP1eqbwe3001},  we obtain
\begin{equation}
P(x,t)\sim \frac{1}{\sqrt{2\pi|x|}}\exp(-t\mathcal{I}_t(l)),
\end{equation}
where
\begin{equation}
\mathcal{I}_t(l)=l\ln(\frac{\sqrt{4pq+l^2}+l}{2pe})-\frac{4pq}{\sqrt{4pq+l^2}+l}+1
\end{equation}
denotes the time rate function.

\subsection{General waiting time PDFs}
In this subsection, we investigate the far tails of the positional distribution and rate functions within the context of a general waiting time PDF. For a general waiting-time distribution, the method in Eq.~\eqref{adlsew102} is no longer applicable. Instead, we resort to the integral representation based on subordination.

Below, we present a method to estimate $f(x|N)$ for large $N$. Note that $p$ and $q$ are independent of the site $j=\cdots, -2, -1, 0, 1, \cdots$.
The probability of arriving at site $x$ after $N$ steps, denoted  as $f(x|N)$, follows
\begin{equation}\label{ATAPPERW1001}
    f(x|N)=\frac{N!}{(\frac{N+x}{2})!(\frac{N-x}{2})!}p^{\frac{N+x}{2}}q^{\frac{N-x}{2}},~~ |x|\leq N.
\end{equation}
It indicates that the particles cannot reach sites larger than $N$ as the length of the displacement is one and the number of steps is $N$.  Moreover, in contrast to continuous displacements, when $N$ is even, the probability of reaching an odd-valued site $x$ is zero; similarly, when $N$ is odd, the probability of reaching an even-valued site is zero.
Using Stirling's approximation $n!\sim\sqrt{2\pi n}(n/e)^n$, Eq.~\eqref{ATAPPERW1001} reduces to
\begin{equation}\label{ATAPPERW1001dp}
  f(x|N)\sim \frac{(2N)^{N+\frac{1}{2}}}{\sqrt{\pi}}\frac{p^{\frac{N+x}{2}}q^{\frac{N-x}{2}}}{(N+x)^{\frac{N+x+1}{2}}(N-x)^{\frac{N-x+1}{2}}}.
\end{equation}
Rewriting Eq.~\eqref{ATAPPERW1001}, we have
\begin{equation}\label{ATAPPERW100g}
\begin{split}
 f(x|N)\sim&\frac{1}{\sqrt{\pi }}2^{N+\frac{1}{2}} N^{N+\frac{1}{2}} (N-x)^{\frac{1}{2} (-N+x-1)}\\
 &\times (N+x)^{\frac{1}{2} (-N-x-1)} p^{\frac{N+x}{2}} q^{\frac{N-x}{2}},
\end{split}
\end{equation}
which will be used to investigate the tails of the distribution of positions.

\subsubsection{Far tails of the distribution}
\begin{figure}[htb]
 \centering
\includegraphics[width=0.5\textwidth]{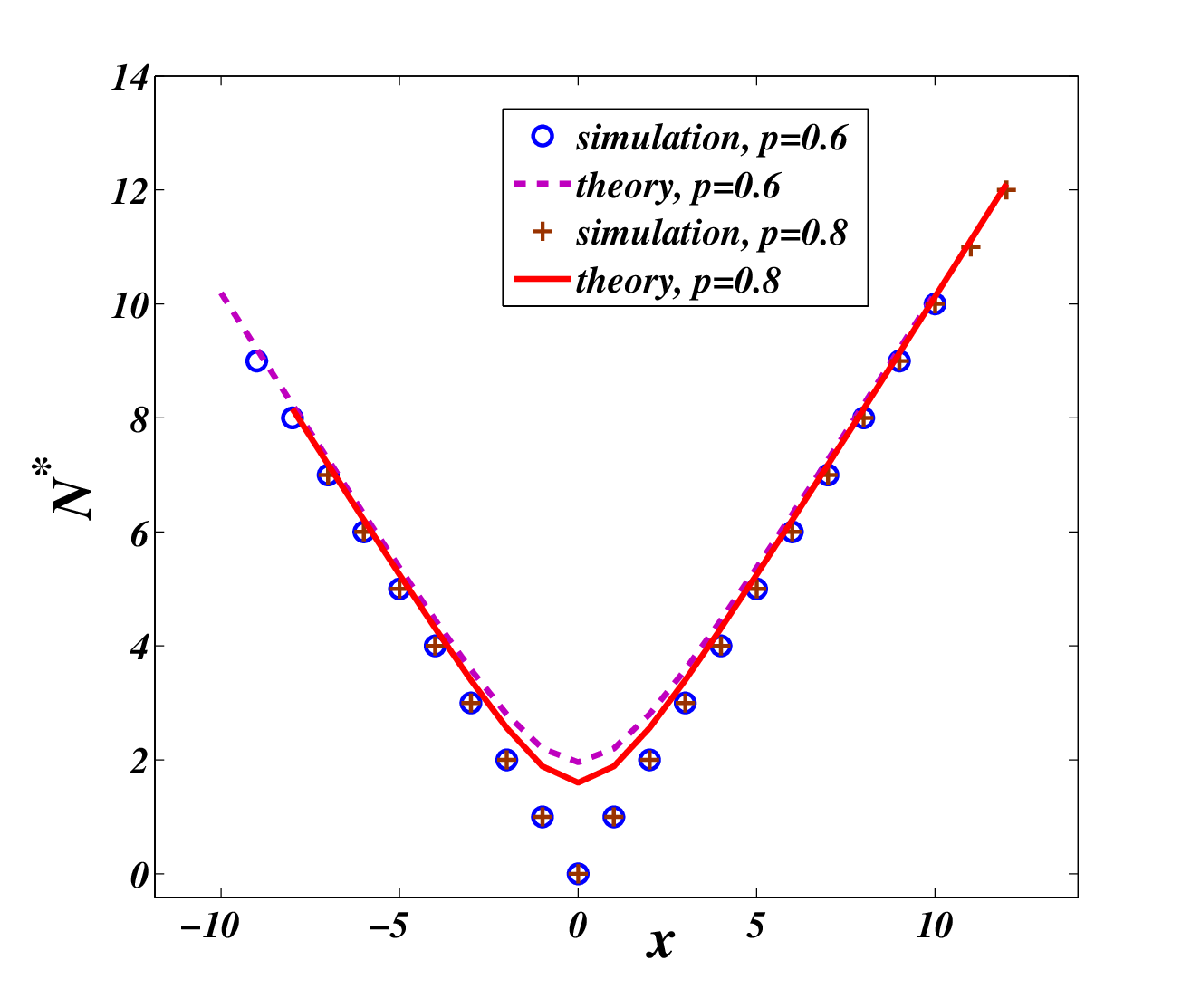}\\
 \caption{ Plot of $N^*$ versus $x$ for various $p$ using Dagum distribution.
The lines represent theoretical predictions, as expressed in Eq.~\eqref{AT19opeq102}, demonstrating a nearly linear increase concerning $x$. Increasing $|x|$ leads to a better convergence.
   }\label{NxingVsXBi}
\end{figure}

From Eq.~\eqref{AT19SEC2EQ1201}, the far tails of the position follow
\begin{widetext}
\begin{equation}\label{AT19opeq100}
P(x,t)\sim \sum_{N=|x|+1}^\infty \exp\left(-\underbrace{\left(\ln\left(\frac{\pi\sqrt {A+1}(N +x)^{\frac {N + x + 1}{2}}(N -
      x)^{\frac {N-x+1} {2}}}{(2 N)^{N} p^{\frac {N + x} {2}}q^{\frac{N - x} {2}}}\right)-\frac{C_{A+1}}{C_A}t-N(A+1)\ln\left(\frac{ezt}{N}\right)\right)}_{\chi(N)}\right).
\end{equation}
\end{widetext}
Our focus is on the statistics for large \( x \). Thus, for \( Q_t(N) \), we apply Eq.~\eqref{StasEli}, while for \( f(x|N) \), we use Eq.~\eqref{ATAPPERW1001dp}. Note that the parameter \( N \) ranges from \( |x|+1 \) to infinity for a given value of \( x \), as Eq.~\eqref{ATAPPERW100g} fails for $N=|x|$.
Next, we apply the saddle-point approximation to estimate the infinite sum in Eq.~\eqref{AT19opeq100}. The first step is to solve for \( N^* \), which satisfies the condition  \( \chi'(N^*) = 0 \), i.e.,
\begin{equation}\label{AT19opeq101}
\frac{\left(4 (N^{*})^2 p q\right) \left(\frac{t z}{N^{*}}\right)^{2 (A+1)}}{(N^{*})^2-x^2}\sim1.
\end{equation}
Note that Eq.~\eqref{AT19opeq101} is solvable for specific exceptional values of \( A \), such as \( A = 1, 2, \dots \), which correspond to cases where the waiting time is analytic at zero.
Below, we focus on the case where $A=0$.
From Eq.~\eqref{AT19opeq101}, the solution is
\begin{equation}\label{AT19opeq102}
N^*\sim\sqrt{4 p q t^2 z^2+x^2}.
\end{equation}
 As expected, $N^{*}$ increases with increasing $x$.
As $x$ becomes sufficiently large, the significance of $p$ diminishes, and we observe that $N^*$ is approximately equal to the absolute value of $x$. This linear relationship is illustrated in Fig.~\ref{NxingVsXBi}. When $A=1$, it can be verified that the leading term coincides with Eq.~\eqref{AT19opeq102}.
Using the saddle point approximation, according to Eqs.~\eqref{AT19opeq100} and \eqref{AT19opeq102}  we find the main result of this section
\begin{equation}\label{AT19opeq103}
\begin{split}
P(x,t)&\sim \frac{\sqrt{\frac{2}{ \left((N^*)^2-x^2\right)}}}{\sqrt{\pi  (A+1)|\chi^{''}(N^*)|}}\exp\Big(
\frac{C_{A+1} }{C_A}t+M[N^*,x,t]\Big)
\end{split}
\end{equation}
with
\begin{equation}
\begin{split}
M[N^*,x,t]=&N^* \ln \left(\frac{\left(2 N^* \sqrt{p q}\right) \left(\frac{e tz}{ N^*}\right)^{A+1}}{\sqrt{(N^*)^2-x^2}}\right)\\
&+\frac{1}{2} x \ln\left(\frac{p (N^*-x)}{q (N^*+x)}\right)
\end{split}
\end{equation}
and
\begin{equation}\label{AT19opeq105}
\begin{split}
\chi^{''}(N^*)\sim &\frac{-(A+1) (N^*)^3+(2 A+1) N^* x^2}{ (N^{*}-x)^2 (N^*+x)^2}\\
&+\frac{-A x^4+(N^{*})^3+N^* x^2}{N^{*} (N^{*}-x)^2 (N^*+x)^2}.
\end{split}
\end{equation}
Recall that $N^*$ is defined in  Eq.~\eqref{AT19opeq102}. It can be seen that the term $M[N^*,x,t]$ is responsible for the exponential decay of the positional distribution.
In addition, when $p=q=1/2$, the position distribution exhibits symmetric far tails.
For this special case, $M[N^*,x,t]$ reduces to
\begin{equation}
\begin{split}
M[N^*,x,t]=&N^* \ln \left(\frac{\left(N^*\right) \left(\frac{e tz}{ N^*}\right)^{A+1}}{\sqrt{(N^*)^2-x^2}}\right)\\
&+\frac{1}{2} x \ln\left(\frac{ N^*-x}{ N^*+x}\right).
\end{split}
\end{equation}
As illustrated in Fig.~\ref{BinomialPxt}, the far tails are sensitive to the bias index $p$. For discussions of the rate function, see Appendix \ref{BBff}.

\begin{figure}[htb]
 \centering
 \includegraphics[width=0.5\textwidth]{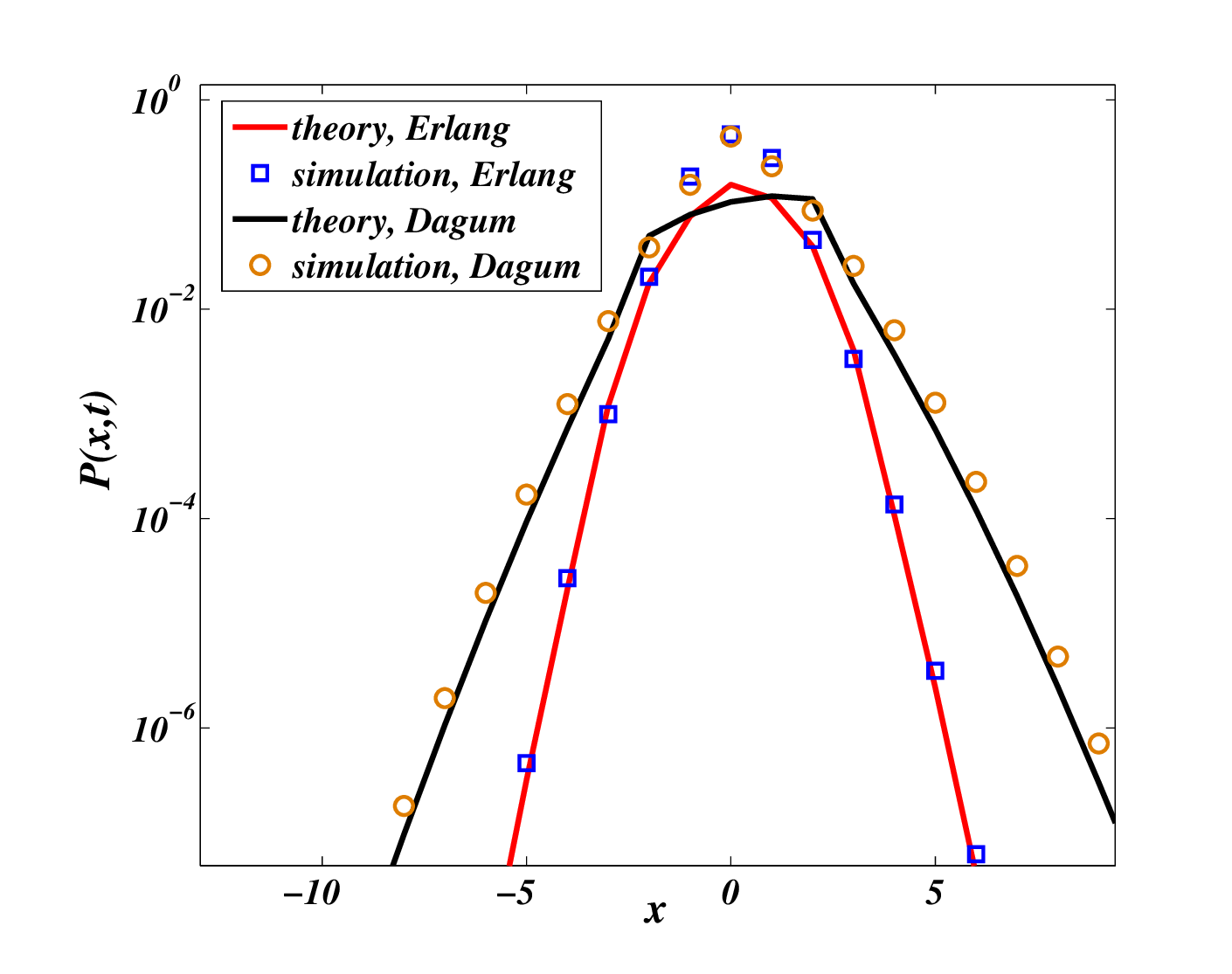}\\
 \caption{ The function $P(x,t)$  for cases where the jump lengths follow a discrete PDF described by Eq.~\eqref{AT19APP4eq102} with different waiting-times PDFs with $p=0.6$, is shown.
  The theoretical approximation, i.e., Eq.~\eqref{AT19opeq103}, is described by the solid lines. The corresponding simulations are plotted by the symbols, i.e.,  `$\Box$' for  Eq.~\eqref{AT19SEC2EQ02} and `$\circ$' for $\phi(\tau)=\tau^{m-1}\exp(-\tau)/(m-1)!$ with $m=2$.
}\label{BinomialPxt}
\end{figure}

%

\section{Conclusion and discussions}
In this manuscript, we investigate the emergence of exponential tails in the context of a biased CTRW model. By combining a large-deviation description of the number of renewals with saddle-point methods, we derive analytical expressions for the far tails of the positional distribution for both continuous (Gaussian) and discrete (two-point) displacement models. The analysis demonstrates that, when the waiting-time distribution is analytic at short times, the exponential decay of the positional distribution is a universal feature, independent of the tails of the waiting-times distribution. Specifically, the behavior is governed by the small $\tau$ expansion of the waiting-time distribution. Moreover,  our method is valid for a family of waiting-time PDFs within a unified framework, without analyzing each case separately.

For Gaussian displacements, we derived the asymptotic expressions for the positional distribution, showing an asymmetric exponential structure induced by the bias. The saddle point analysis gives a theoretical prediction for the most likely number of steps reaching the position $x$. Furthermore, an Einstein-like relation between biased and unbiased propagators was established in the far-tail regime. For weak bias, this relation reduces to a simple exponential form, while higher-order corrections become relevant for stronger bias. We further extended the theory to the binomial random walk using discrete displacements. Despite the  constraint between the number of jumps and the position, the far tails exhibit a nearly exponential decay at the short time limit. The analysis reveals that the dominant contribution arises from trajectories with a number of steps scaling linearly with the displacement. 
Although $N^*$ is insensitive to the value of $p$, the tails of the positional distribution depend strongly on $p$. This result provides a nontrivial extension of previous studies \cite{Burov2022Exponential}, where discrete jump processes were not explicitly addressed. In addition, for both cases, two types of rate functions, associated with position and time, were introduced and analyzed. These rate functions quantify the probabilities of rare events and are directly responsible for the nearly exponential decay observed in the positional distributions.

Overall, the findings of this work provide further insight into the exponential decay in biased stochastic transport processes and highlight the role of the Taylor expansion of the waiting-time distribution in shaping the far tails of positional distributions. Given that the biased CTRW model is widely used to describe anomalous chemical transport in porous media \cite{Alon2017Time} or other complex systems, these results are expected to provide useful guidance for experimental observations of exponential decay.

Throughout the manuscript, we restrict ourselves to waiting-time distributions that are analytic at small $\tau$, as given by Eq.~\eqref{AT19SEC2EQ101}. A natural question is what changes when the waiting times are instead drawn from the one-sided L{\'e}vy distribution. At least according to numerical simulations, the far tails of the positional distribution decay faster than an exponential function. For the non-analytic case, the rate-function approach developed by Sokolov and Pacheco \cite{Adrian2021Large} may offer advantages over saddle-point methods. Furthermore, it is worth examining the convergence problem for the aforementioned two methods, along with the role of aging effects \cite{Barkai2003Aging,Allegrini2003Generalized,Schulz2014Aging,Wang2017Aging}.

\begin{acknowledgments}
W.W. thanks Eli Barkai and Stanislav Burov for their insightful discussions and for motivating this work. W.W. was supported by the National Natural Science
Foundation of China under Grant No. 12105243 and the Zhejiang Province Natural Science Foundation LQ22A050002.

\end{acknowledgments}
\appendix
\begin{appendices}
\begin{figure}[htb]
 \centering
\includegraphics[width=0.5\textwidth]{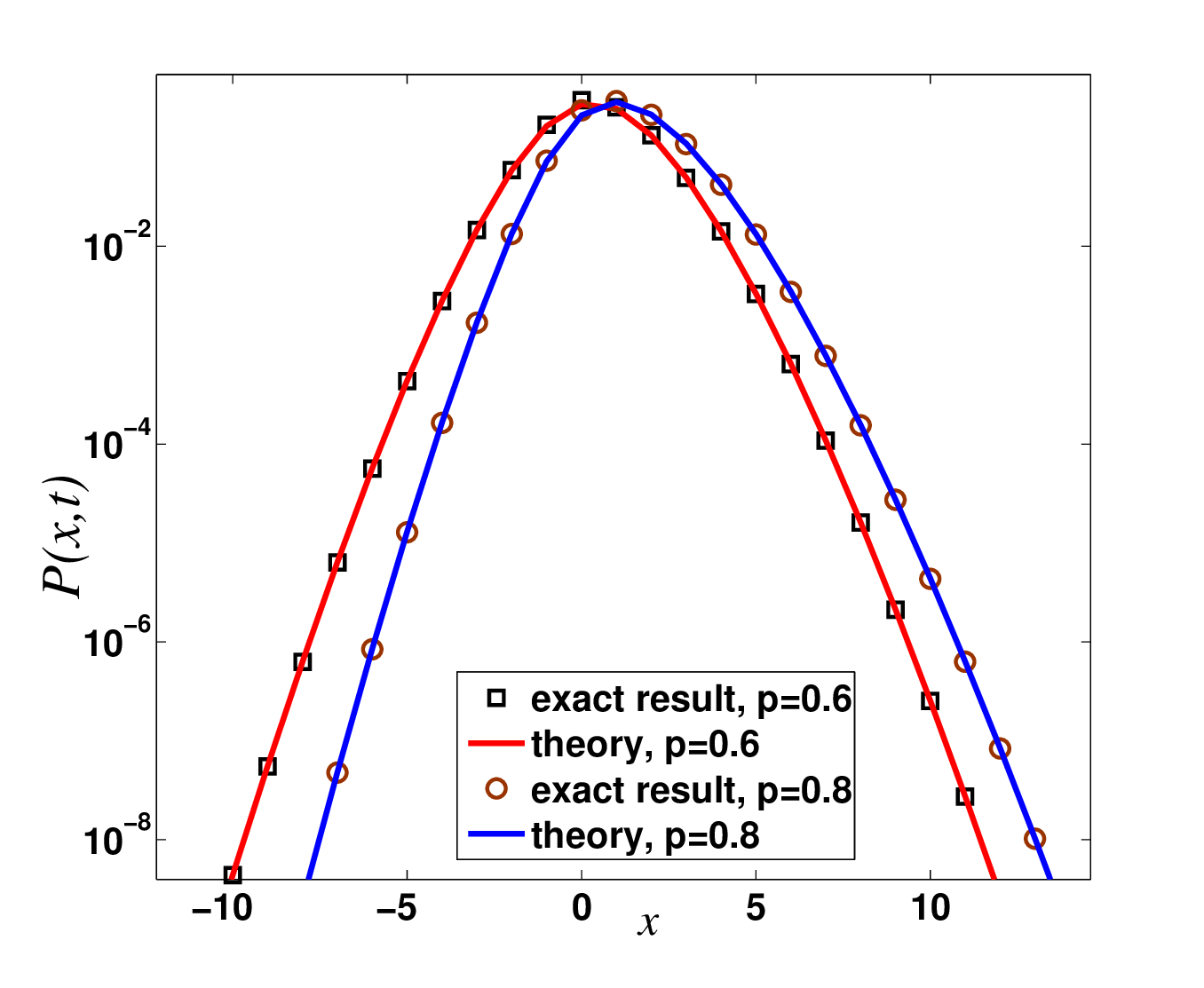}\\
\caption{ Decay of the positional distribution, where waiting times are drawn from the exponential distribution $\phi(\tau)=\exp(-\tau)$ and the displacements are generated from Eq.~\eqref{AT19APP4eq102}.  The symbols describe the exact result for various $p$, obtained from Eq.~\eqref{AT19SEC2EQ1201}, and the solid lines are the corresponding theoretical prediction Eq.~\eqref{AT19APP1eqbwe3001}.  The observation time is $t=2$.
 }\label{ExpPxt}
\end{figure}

\begin{figure}[h]
 \centering
 \includegraphics[width=0.50\textwidth]{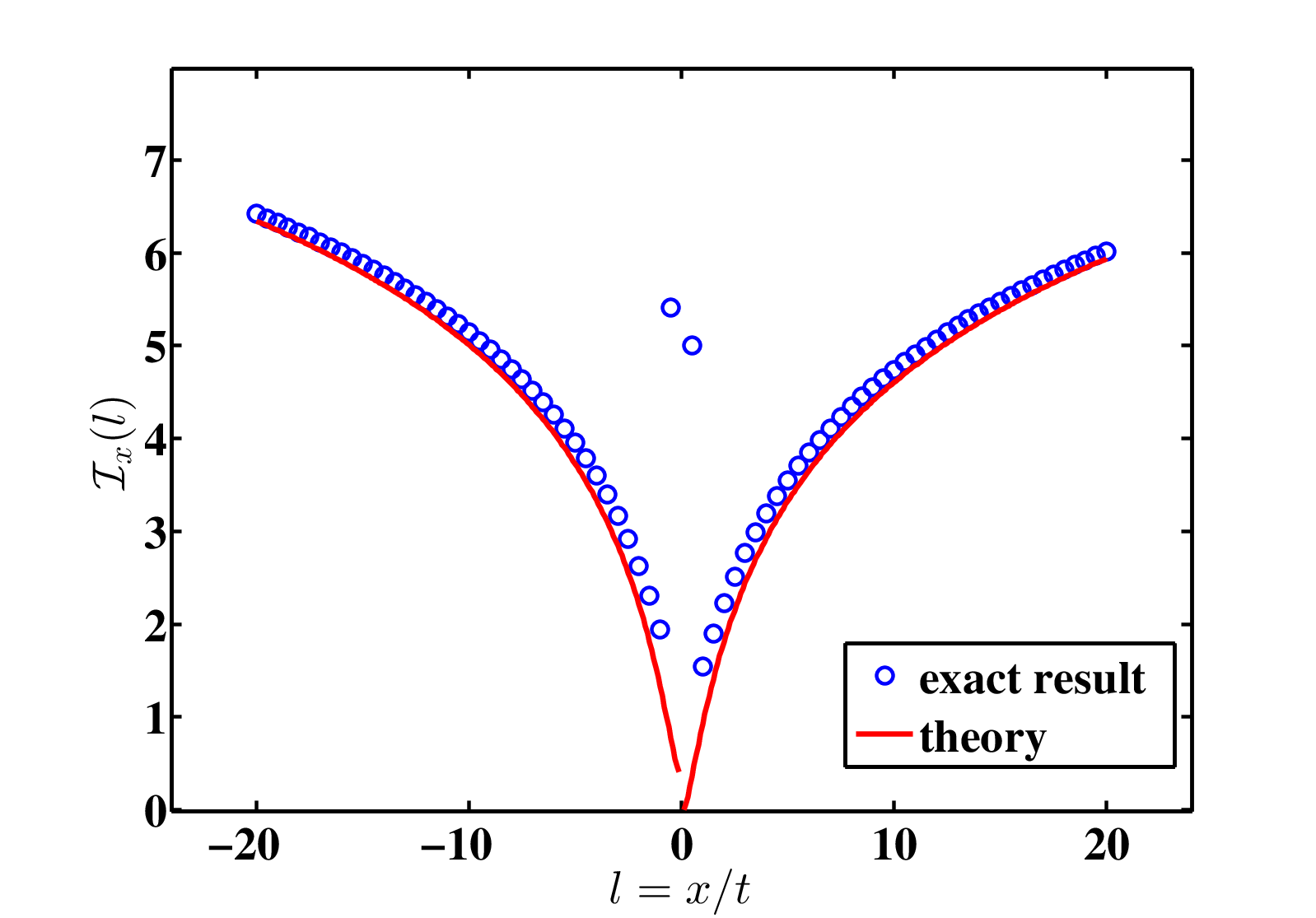}\\
 \caption{Rate function $\mathcal{I}_x(l)=|l|\mathcal{I}_t(l)$, where the waiting time follows Erlang distribution $\phi(\tau)=\tau^{m-1}\exp(-\tau)/(m-1)!$ with $m=2$. Here the solid line shows the theoretical prediction $|l|\mathcal{I}_t(l)$ with $\mathcal{I}_t(l)$ obtained from Eq.~\eqref{RateBioulTime} and the related exact results are obtained from Eqs.~\eqref{AT19SEC2EQ1201} and \eqref{ATAPPERW1001} with $Q_t(N)=(\Gamma (2N+2,t)/\Gamma (2N+2)-\Gamma (2N,t)/\Gamma (2N)$ \cite{Wang2020Large}, where $\Gamma(x,y)$  is the  incomplete  gamma function \cite{Abramowitz1984Handbook}.
 The parameters are $t=2$, $p=0.6$ and $q=0.4$.
}\label{RateFunctionErlangM2}
\end{figure}

\begin{figure}[htb]
 \centering
 \includegraphics[width=0.5\textwidth]{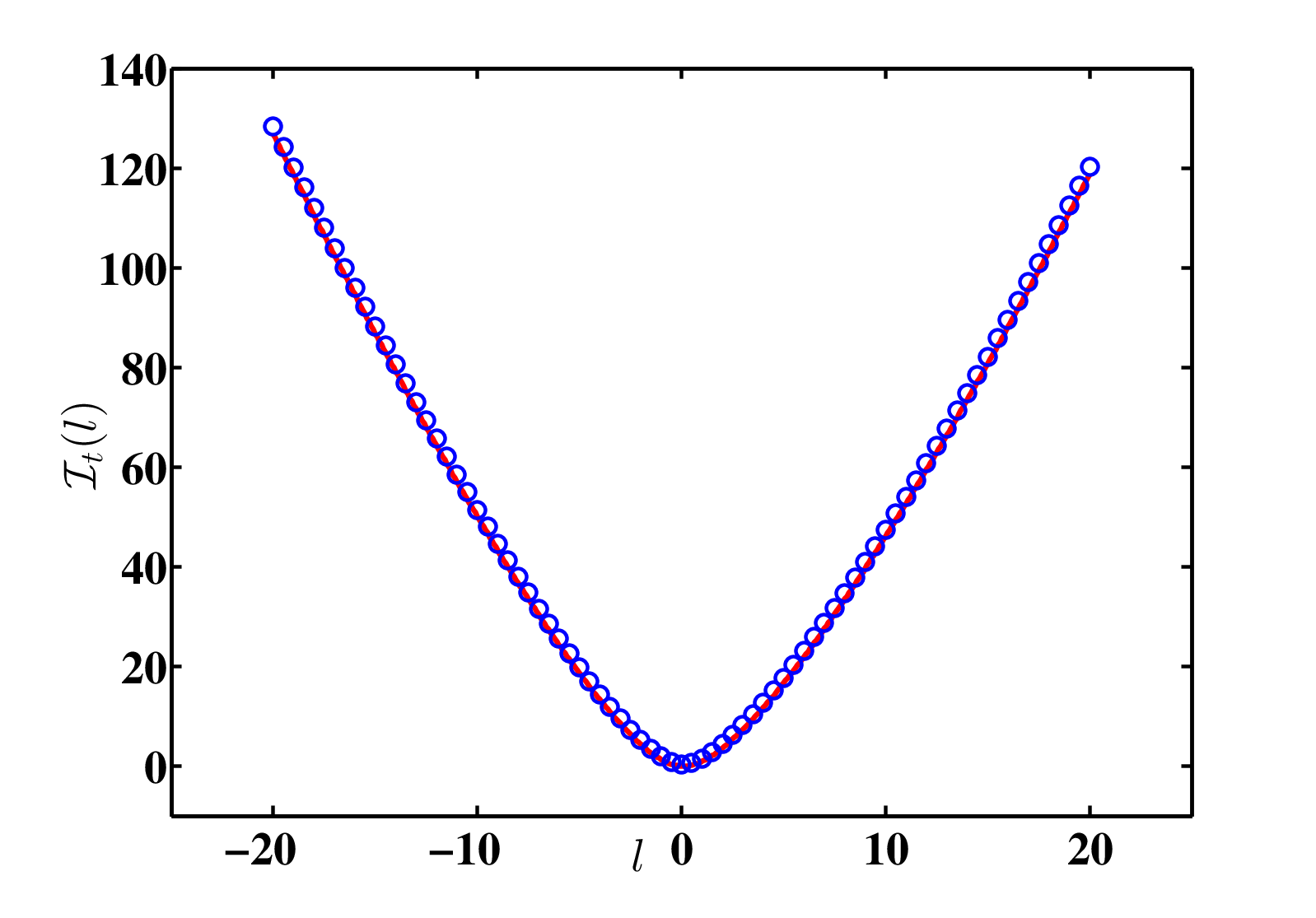}\\
 \caption{Time Rate function $\mathcal{I}_t(l)$.
 The solid line shows the plot of Eq.~\eqref{RateBioulTime}, and the symbols illustrate the exact results.
 The parameters are the same as in Fig.~\ref{RateFunctionErlangM2}.
}\label{RateFunctionTimeErlangm2}
\end{figure}

\section{The waiting-time distributions used in the simulation}
In our simulations, we use a well-known waiting-time PDF called the exponential distribution
\begin{equation}\label{expPDF}
\phi(\tau)=\exp(-\tau),
\end{equation}
where waiting times $\tau$ have a finite mean and variance. According to Eq.~\eqref{AT19SEC2EQ101}, we have $A=0$, $C_0=1$, and $C_1=-1$.

The second distribution is a broad distribution with infinite mean and variance. Here we consider a special case of the power-law distribution, namely the Dagum distribution
\begin{equation}\label{AT19SEC2EQ02}
\phi(\tau)=\frac{\tau_0}{(\tau_0+\tau)^2}
\end{equation}
with $\tau_0>0$. For our simulations, we choose $\tau_0=1$.
When $\tau \to 0$, the Taylor expansion of the Dagum distribution yields $\phi(\tau) \sim 1 - 2\tau$.
Therefore, from Eq.~\eqref{AT19SEC2EQ101}, it follows that $A = 0$, $C_A = 1$, and $C_{A+1} = -2$.

The third waiting-time PDF is the Erlang distribution \cite{Forbes2011Statistical}. It corresponds to the distribution of the sum of $m$ independent exponential random variables, each with mean $1$. In real space, it reads
\begin{equation}
\phi(\tau)=\frac{\tau^{m-1} e^{-\tau}}{(m-1)!}.
\end{equation}
When $\tau\to 0$, one has the asymptotic behavior
$\phi(\tau)\sim (\tau^{m-1}-\tau^{m})/(m-1)!$.

\section{The limit of Rate function for $l\to 0$}

Here we discuss the limit of the position rate function for  Gaussian displacements. For $|l|\to 0$, Eq.~\eqref{AT19SEC3fewsis1001add11} simplifies to
\begin{equation}\label{AT19SEC3fsis1006}
\begin{split}
\mathcal{ I}_x(l)&=-\frac{\sqrt{A+1} }{ \sigma}\frac{1}{\sqrt{g l^2}}+\frac{\sqrt{A+1} \sqrt{g l^2}}{ \sigma}\\
&~~~~~~-\frac{a}{\sigma^2}\sign(l)-\frac{C_{A+1}}{C_A}\frac{1}{|l|}.
\end{split}
\end{equation}
Combining Eqs.~\eqref{AT19SEC3fsds101} and \eqref{AT19SEC3fsis1006}, we have
\begin{equation}\label{AT19SEC3fsis1007}
\begin{split}
P(x,t)\simeq &\exp \Big(\frac{a }{\sigma ^2}x+\frac{\sqrt{A+1} }{\sqrt{g} \sigma }t-\frac{\sqrt{A+1} \sqrt{g} }{(2 \sigma )}\frac{x^2}{t}\\
&~~~+\frac{C_{A+1}}{C_A} t\Big).
\end{split}
\end{equation}
The above form corresponds to the Gaussian law, consistent with the central limit theorem.

\section{Rate function for discrete case}\label{BBff}
Now, we consider the corresponding rate functions, namely the position and the time rate functions, which describe the limiting law for large $x$ and large $t$, respectively.
Rewriting Eq.~\eqref{AT19opeq103}, we have
\begin{equation}
\begin{split}
-\ln(P(x,t))&\sim -t\mathcal{I}_t(l)
\end{split}
\end{equation}
with $l=x/t$  and
\begin{equation}\label{RateBioulTime}
\begin{split}
\mathcal{I}_t(l)=&-\frac{C_{A+1}}{C_A}-\frac{1}{2} l \ln \left(\frac{p \left(\sqrt{l^2+4 p q z^2}-l\right)}{q \left(\sqrt{l^2+4 p q z^2}+l\right)}\right)\\
&-\sqrt{l^2+4 p q z^2} \\
&\times\ln \left(\sqrt{\frac{l^2}{z^2}+4 p q} \left(\frac{e z}{\sqrt{l^2+4 p q z^2}}\right)^{A+1}\right).
\end{split}
\end{equation}
Using the relation $\mathcal{I}_t(l)=|l|\mathcal{I}_x(t)$ and Eq.~\eqref{RateBioulTime}, the position rate function can be obtained.
\section{Figures for simulations together with theoretical predictions}\label{CCCCC}
Here, we present additional figures to further validate the theoretical predictions. In Fig.~\ref{ExpPxt}, we verify Eq.~\eqref{AT19APP1eqbwe3001} for an exponential waiting-time distribution. In Figs.~\ref{RateFunctionErlangM2} and \ref{RateFunctionTimeErlangm2}, we display the corresponding position rate function and time rate function for discrete displacements, respectively.

\end{appendices}

\bibliographystyle{prestyle}
\bibliography{wenxian}

\end{document}